\documentclass[aps,pra,showpacs,reprint,twocolumn]{revtex4-1}
\usepackage{amsfonts}
\usepackage{amsmath}

\begin{document}

\title{Comment on ``Connection between entanglement and the speed of quantum
 evolution''}

\author{H.~F. Chau}\email{hfchau@hkusua.hku.hk}
\affiliation{Department of Physics and Center of Computational and Theoretical
 Physics, University of Hong Kong, Pokfulam Road, Hong Kong}
\date{\today}

\begin{abstract}
 Batle \emph{et al.} [Phys. Rev. A {\bf 72}, 032337 (2005)] and Borr\'{a}s
 \emph{et al.} [Phys. Rev. A {\bf 74}, 022326 (2006)] studied the connection
 between entanglement and speed of quantum evolution for certain
 low-dimensional bipartite quantum states.  However, their studies did not
 cover all possible cases.  And the relation between entanglement and the
 maximum possible quantum evolution speed for these uncovered cases can very
 different from the ones that they have studied.
\end{abstract}


\maketitle

 Batle \emph{et al.}~\cite{ent1} studied those pure two-qubit states that can
 evolve to their orthogonal subspaces under the action of the local
 time-independent Hamiltonian $H_A \otimes H_B$ in which the spectra of the
 two-level Hamiltonians $H_A$ and $H_B$ equal $\{ 0, \epsilon \}$ for some
 $\epsilon > 0$.  They denoted the energy eigenstates of $H_A$ and $H_B$ by
 $|0\rangle$ and $|1\rangle$ so that $H_i |0\rangle = 0$ and $H_i |1\rangle =
 \epsilon |1\rangle$ for $i = A, B$.  By writing a normalized pure two-qubit
 state $|\psi\rangle$ of two distinguishable particles in the form $c_0
 |00\rangle + c_1 |01\rangle + c_2 |10\rangle + c_3 |11\rangle$, Batle
 \emph{et al.} deduced that the time $t$ at which $|\psi\rangle$ evolves to
 its orthogonal subspace is given by the quadratic equation~\cite{ent1}
\begin{equation}
 |c_0|^2 + ( |c_1|^2 + |c_2|^2 ) z + |c_3|^2 z^2 = 0 ,
 \label{E:quad}
\end{equation}
 where $z = \exp ( i \epsilon t / \hbar )$.  In Ref.~\cite{ent1}, Batle
 \emph{et al.} were interested in those states that must evolve to their
 orthogonal subspaces at some time.

 Note that there are two cases to consider, namely, the generic case in which
 the leading coefficient of the above quadratic equation, $|c_3|^2$, is
 non-zero and the singular case in which $|c_3|^2 = 0$.

 Batle \emph{et al.} only considered the generic case in Ref.~\cite{ent1}.
 In this case, the condition that $|\psi\rangle$ must evolve to its orthogonal
 subspace at some time implies $|c_0|^2 = |c_3|^2 > 1/4$.  By solving
 Eq.~(\ref{E:quad}), Batle \emph{et al.} obtained an expression for the time
 $\tau$ when $|\psi\rangle$ first evolved to its orthogonal subspace.  Combined
 with the literature results that $\tau$ is lower-bounded by
\begin{equation}
 T_\text{min} \equiv \min \left( \frac{\pi\hbar}{2E} ,
 \frac{\pi\hbar}{2\Delta E} \right)
 \label{E:T_min}
\end{equation}
 where $E$ and $\Delta E$ are the average energy and the standard deviation of
 the energy of the state, respectively, Batle \emph{et al.} showed
 that~\cite{ent1}
\begin{equation}
 \tau \geq T_\text{min} = T_\text{min1} \equiv \frac{\pi\hbar}{2\sqrt{2}
 \epsilon |c_0|} = \frac{\pi\hbar}{2\sqrt{2} \epsilon |c_3|} .
 \label{E:T_min1}
\end{equation}
 Since the concurrence $C$ of the state is given by the equation
\begin{equation}
 C^2 = 4 \left| |c_0|^2 - e^{i\phi} \sqrt{\delta ( 1 - \delta )} 2 |c_0|^2 \cos
 \alpha \right|^2
 \label{E:concurrence_relation}
\end{equation}
 for some parameters $\phi\in {\mathbb R}$ and $\delta\in [0,1]$.  By means of
 the observation that $|c_0|^2 \leq ( 1 + |C| ) / 4$ for a fixed concurrence
 $C$, they deduced from Eqs.~(\ref{E:T_min1})
 and~(\ref{E:concurrence_relation}) that
\begin{equation}
 \frac{\tau}{T_\text{min}} \geq \frac{\sqrt{2 ( 1 + |C| )}}{\pi} \cos^{-1}
 \left( \frac{|C|-1}{|C|+1} \right) \geq 1 .
 \label{E:tau_bound}
\end{equation}
 Most importantly, they concluded from Eq.~(\ref{E:tau_bound}) that $\tau =
 T_\text{min}$ if and only if the state was maximally entangled~\cite{ent1}.

 Batle \emph{et al.}, however, did not consider the singular case in which
 $|c_3|^2 = 0$ in Ref.~\cite{ent1}.  For the singular case, Eq.~(\ref{E:quad})
 becomes a linear equation.  (Should $|c_3|^2 = |c_1|^2 + |c_2|^2 = 0$ so that
 L.H.S. of Eq.~(\ref{E:quad}) becomes a constant, the state $|\psi\rangle$ can
 never evolve to its orthogonal subspace.  So, the case in which
 Eq.~(\ref{E:quad}) is a linear equation is the only unanalyzed situation.)
 The condition that the state $|\psi\rangle$ with $|c_3|^2 = 0$ can evolve to
 its orthogonal subspace is $|c_0|^2 = |c_1|^2 + |c_2|^2 = 1/2$.  And by
 solving Eq.~(\ref{E:quad}), we arrive at $\tau = \pi\hbar / \epsilon =
 \pi\hbar / 2E = \pi\hbar / 2 \Delta E = T_\text{min}$.  Most importantly,
 $\tau$ cannot be expressed as a function of the concurrence $C = 2 |c_1|
 |c_2|$ of the state and $T_\text{min1}$ is not well-defined as $|c_0|^2 \neq
 |c_3|^2$.  In fact, for each $C$, there is a state that attains the least
 evolution time $T_\text{min}$.  An example is the family of states $(
 |00\rangle + \sqrt{x} |01\rangle + \sqrt{1-x} |10\rangle ) / \sqrt{2}$ for $x
 \in [0,1]$ whose concurrence $C = 2 \sqrt{x(1-x)}$.  This family of states
 violates the first inequality in Eq.~(\ref{E:tau_bound}) provided that $0 < x
 < 1$.  It also shows that partially entangled states can attain the evolution
 time lower bound $T_\text{min}$.  Thus, the relation between entanglement and
 the time needed to evolve to the orthogonal subspace for the singular case can
 be very different from the generic case.
 
 Since Batle \emph{et al.} did not discuss similar singular cases for bosonic
 two-qubit pure states and fermionic two-qutrit pure states of
 indistinguishable particles in Ref.~\cite{ent1}, their analysis in these two
 situations are also incomplete.  In fact, the case of $x = 1/2$ in the above
 family of states is an example of a non-maximally entangled bosonic two-qubit
 pure state with the least possible evolution time.  Their followup
 paper~\cite{ent2} dealing with the extensions to the cases of mixed states and
 evolution into non-orthogonal states are also incomplete as it suffers the
 same problem of effectively restricting the analysis only to generic
 situations because they sample the initial states according to the Haar
 measure in their Monte Carlo simulations.

\begin{acknowledgments}
 I would like to thank F.~K. Chow, C.-H.~F. Fung and K.~Y. Lee for their
 discussions.
 This work is supported by the RGC grant number HKU~700709P of the HKSAR
 Government.
\end{acknowledgments}

\bibliography{qc49.3}
\end{document}